\documentclass[prd,aps,secnumarabic,showpacs]{revtex4}
\begin{document}
\title{Muon decay in a laser field}
\renewcommand{\thefootnote}{\fnsymbol{footnote}}
\author{Duane A. Dicus$^{1,}$\footnote{Electronic address: dicus@physics.utexas.edu},
Arsham Farzinnia$^{2,}$\footnote{Electronic address: farzinni@msu.edu},
Wayne W. Repko$^{2,}$\footnote{Electronic address: repko@pa.msu.edu},
and Todd M. Tinsley$^{3,}$\footnote{Electronic address: tinsley@hendrix.edu}}
\affiliation{$^1$Physics Department, University of Texas, Austin, TX 78712  \\
$^2$Department of Physics and Astronomy, Michigan State University, East Lansing MI 48824  \\
$^3$Department of Physics, Hendrix College, Conway AR 72034}
\renewcommand{\thefootnote}{\arabic{footnote}}

\date{\today}

\begin{abstract}
We investigate the change in the decay rate of a muon caused by embedding it in the field of a laser.
A previous paper found that the change could be large, as much as an order of magnitude.  We find the more
intuitive result that the change is small and give analytic expressions for the small corrections.
\end{abstract}
\maketitle

\section{Introduction}

There is interest and work on the properties of elementary systems when they are placed in strong
electromagnetic fields\cite{intro,Todd,DRT}.  Recently, in an attempt to extend this work to unstable systems, Liu, Li, and
Berakdar\cite{LLB} (LLB) calculated the effect of a strong laser field on the decay rate of muons.  They found
the lifetime could be changed from its normal value of $2.2\times10^{-6}$ seconds to $5\times10^{-7}$ seconds or
even less.
This conclusion was challenged by Narozhny and Fedotov, who offer an abbreviated calculation to support their
criticism \cite{rus,LLB2}.
If LLB were correct, this would be a very interesting result.
We have done our own calculation and, unfortunately, also reach a very different conclusion. We find the effects
of the laser to be very small and give explicit expressions for these small effects.

Although the idea of the problem is straightforward, the actual calculation is somewhat complicated and LLB did
the complicated part numerically.  We do everything analytically.  Because of this difference,
and because we get such a different result, we will present our
calculation in some detail and only after our results are apparent will we compare with LLB.
Also because our calculation is analytic, we don't need to make definite choices about the properties
of the laser.  We will assume only that the energies of the photons are about $0.1-1.0$\,eV and that the magnitude of
the laser field amplitude is $10^6-10^7 {\rm V/cm}$ (as used in LLB).
The next section gives our work, the following section compares with, and discusses,
LLB,
including the fact that we have somewhat different starting points; we use the Volkov wavefunction \cite{V1,V2,BLP,SVTM}
for circular polarization while they use an approximation to the wavefunction for linear polarization.
The last section repeats our conclusions.

\section{Formalism}

The process is muon decay,
\begin{equation}\label{decay}
\mu^-(P)\,\longrightarrow\,e^-(p)+\bar{\nu_e}(q_1)+\nu_{\mu}(q_2)\,,
\end{equation}
where the arguments are our labels for the momentum.
We will assume the photons from the laser are along the z-axis with momentum $k^{\mu}\,=\,(\omega,0,0,\omega)$
and circular polarization,
\begin{eqnarray}
A^{\mu}(x)\,&=&\,a\,n_1^{\mu}\cos\,k\cdot\,x+a\,n_2^{\mu}\sin\,k\cdot\,x\,,  \\
n_1^{\mu}\,&=&\,(0,1,0,0)\,,  \\
n_2^{\mu}\,&=&\,(0,0,1,0)\,.
\end{eqnarray}
The electron wave function is then
\begin{equation}\label{wf}
\psi(x)\,=\,e^{-i\frac{ea}{p\cdot\,k}p_x\sin\,k\cdot\,x}e^{-i\frac{e^2a^2}{2p\cdot\,k}k\cdot\,x-ip\cdot\,x}
            \Big(1+\frac{ek\!\!\!/A\!\!\!/}{2p\cdot\,k}\Big)\,u(p)
\end{equation}
where we have taken the electron to be in the $xz$ plane and
thereby avoided a factor
$-i\frac{ea}{p\cdot\,k}p_y\,\cos\,k\cdot\,x$ in the
exponential. From the second exponential factor the electron
has an effective momentum and mass
\begin{eqnarray}
q^{\mu}\,&=&\,p^{\mu}+\frac{e^2a^2}{2p\cdot\,k}k^{\mu}\label{q}\,,  \\
m^2\,&=&\,m_0^2+e^2a^2\,,
\end{eqnarray}
where $m_0=0.511\,{\rm MeV}$.  Note that $q\cdot\,k\,=\,p\cdot\,k$ and $q_x\,=\,p_x$.
Following the standard proceedure we use the generating function for Bessel functions\cite{AS},
\begin{equation}\label{gf}
e^{\frac{1}{2}z(t-1/t)}\,=\,\sum_{\ell=-\infty}^{\infty}t^{\ell}J_{\ell}(z)
\end{equation}
to rewrite the first factor in (\ref{wf}) as
\begin{equation}\label{gfwf}
e^{-i\frac{ea}{p\cdot\,k}p_x\sin\,k\cdot\,x}\,=\,\sum_{\ell=-\infty}^{\infty}J_{\ell}(D)e^{-i\ell\,k\cdot\,x}
\end{equation}
with $D\,=\,\frac{eap_x}{p\cdot\,k}$.
Momentum conservation is then
\begin{equation}\label{CM}
P^{\mu}+\ell\,k^{\mu}\,=\,q^{\mu}+q_1^{\mu}+q_2^{\mu}
\end{equation}
and the matrix element, for a given value of $\ell$, is
\begin{equation}\label{T}
R_{\ell}\,=\,\frac{G}{\sqrt{2}}\bar{u}(q_2)\gamma^{\alpha}(1-\gamma_5)u(P)\,\bar{u}(p)\big[\Delta_0+\Delta_1n\!\!\!/_1k\!\!\!/
                                                     +\Delta_2n\!\!\!/_2k\!\!\!/\big]\gamma_{\alpha}(1-\gamma_5)v(q_1)
\end{equation}
where
\begin{eqnarray}
\Delta_0\,&=&\,J_{\ell}(D)  \\
\Delta_1\,&=&\,\frac{1}{2}\frac{ea}{2p\cdot\,k}\big(J_{\ell+1}(D)+J_{\ell-1}(D)\big)   \\
\Delta_2\,&=&\,\frac{-i}{2}\frac{ea}{2p\cdot\,k}\big(J_{\ell+1}(D)-J_{\ell-1}(D)\big)\,.
\end{eqnarray}
Note that the argument of the electron spinor is still $p$.

We square the matrix element in the usual way, using FORM\cite{form}, and integrate out the momentum of the neutrinos
in the usual way using
\begin{equation}\label{neuint}
\int\frac{d^3q_1}{2q_1^0}\frac{d^3q_2}{2q_2^0}\delta^{4}(Q-q_1-q_2)q_1^{\alpha}q_2^{\beta}\,=\,
    \frac{\pi}{24}\big(Q^2\,g^{\alpha\beta}+2Q^{\alpha}Q^{\beta}\big)\Theta(Q^2)\,.
\end{equation}
The total width is then
\begin{equation}\label{TG}
\Gamma\,=\,\sum_{\ell=-\infty}^{\infty}\Gamma_{\ell}
\end{equation}
with the width for each $\ell$ given as an integral over the electron energy and angle
\begin{equation}\label{Gl}
\Gamma_{\ell}\,=\,\frac{1}{3072\pi^3M}\int\,dE\,|{\bf q}|\int\,dz\,\Theta(Q^2)\,|T_{\ell}|^2
\end{equation}
where $E$ is $q^0$ (the Jacobian connecting $p^{\mu}$ and $q^{\mu}$ is unity) and $M$ is the muon mass.

\newpage

The square of the matrix element, (\ref{T}), summed over spin and integrated over the neutrino momentum, is
\begin{eqnarray}
|T_{\ell}|^2\,&=&\,128G^2\,\Big\{J_{\ell}^2(D)\big[3P\cdot\,q(M^2+m^2)-4(P\cdot\,q)^2-2M^2m^2\big]  \nonumber \\
              &+&\,J_{\ell}^2(D)\frac{e^2a^2}{2q\cdot\,k}\big[2q\cdot\,k(M^2-P\cdot\,q)
                                                           +P\cdot\,k(4P\cdot\,q-3M^2-m^2)\big] \nonumber \\
              &+&\,\ell\,J_{\ell}^2(D)\big[q\cdot\,k(2M^2-4P\cdot\,q)-P\cdot\,k(3M^2-8P\cdot\,q+3m^2)\big] \nonumber \\
              &+&\,2\ell\,J_{\ell}^2(D)\frac{e^2a^2}{q\cdot\,k}P\cdot\,k(q\cdot\,k-P\cdot\,k) \nonumber \\
              &+&\,4\ell^2\,J_{\ell}^2(D)P\cdot\,k(q\cdot\,k-P\cdot\,k)  \nonumber  \\
              &+&\,\frac{e^2a^2}{4\,q\cdot\,k}[J_{\ell+1}^2(D)\,+\,J_{\ell-1}^2(D)]
                                 \big[-4P\cdot\,qP\cdot\,k+3M^2P\cdot\,k+m^2P\cdot\,k+2q\cdot\,k(P\cdot\,q-M^2)\big]  \nonumber  \\
              &+&\,\frac{e^2a^2}{q\cdot\,k}\,\ell\,[J_{\ell+1}^2(D)+J_{\ell-1}^2(D)]
                                                                        P\cdot\,k(P\cdot\,k-q\cdot\,k)  \nonumber  \\
              &+&i\epsilon(P,q,k,n_2)\frac{ea}{2q\cdot\,k}J_{\ell}(D)[J_{\ell+1}(D)-J_{\ell-1}(D)]
                                                                                      (3M^2+m^2-4P\cdot\,q)  \nonumber  \\
              &+&i\epsilon(P,q,k,n_2)\frac{ea}{q\cdot\,k}\ell\,J_{\ell}(D)[J_{\ell+1}(D)-J_{\ell-1}(D)]
                                                                                      (2P\cdot\,k-q\cdot\,k)   \nonumber  \\
              &+&i\epsilon(P,k,n_1,n_2)\,\frac{e^2a^2}{4\,q\cdot\,k}[J_{\ell+1}^2(D)-J_{\ell-1}^2(D)]
                                                                                      (3M^2+m^2-4P\cdot\,q)  \nonumber  \\
              &+&i\epsilon(P,k,n_1,n_2)\,\frac{e^2a^2}{2\,q\cdot\,k}\ell\,[J_{\ell+1}^2(D)-J_{\ell-1}^2(D)]
                                                                                (2P\cdot\,k-2q\cdot\,k)  \nonumber  \\
              &+&i\epsilon(k,q,n_1,n_2)\,\frac{e^2a^2}{2\,q\cdot\,k}[J_{\ell+1}^2(D)-J_{\ell-1}^2(D)]
                                                              (M^2-P\cdot\,q)\Big\} \label{T2}
\end{eqnarray}
where $\epsilon^{0123}=-i$.
%THIS MINUS CORRECTS FOR WAYNE'S CHANGE OF SIGN COMPARED TO MY NOTES.

If we set $a=0$ then $D=0$ so $J_{\ell}=0$ for $\ell\ne\,0$ and $J_0(0)=1$ and we get the usual expression
for muon decay from the first line of (\ref{T2})
\begin{equation}\label{G0}
\Gamma^0\,=\,\frac{G^2M^5}{192\pi^3}
\end{equation}
where terms proportional to the electron mass have been neglected\footnote{The width including the electron mass terms is known to be
$\Gamma\,=\,\Gamma^0(1-8r^2+8r^6-r^8-12r^4\ln\,r^2)$, $r=\frac{m}{M}$, and comes from the first line of (\ref{T2})
if $J_{\ell}^2(D)$ is set to unity.}.
Note that $\Gamma^0$ is not the same as $\Gamma_{\ell=0}$.

The crucial thing is the limits on the integrals in Eq.\,(\ref{Gl}).  These are determined by the $\Theta$ function,
\begin{eqnarray}
\int\,dE\,\int\,dz\,\Theta(Q^2)\,&=&\,\int_{m}^{\frac{M}{2}}\,dE\,\int_{-1}^1dz\,
       +\,\int_{\frac{M}{2}}^{\frac{M}{2}+\ell\omega}\,dE\,\int_{z_L(E)}^1dz\,,\label{I1}\:\:\:\:\:\:\:(\ell\,>\,0)\label{lim1} \\
\int\,dE\,\int\,dz\,\Theta(Q^2)\,&=&\,\int_{m}^{\frac{M}{2}+\ell\omega}\,dE\,\int_{-1}^1\,dz\,
       +\,\int_{\frac{M}{2}+\ell\omega}^{\frac{M}{2}}\,dE\,\int_{-1}^{z_L(E)}dz\,,\:\:\:\:\:\:\:(\ell\,<\,0)\label{lim2}
\end{eqnarray}
where
\begin{equation}
z_{L}(E)\,=\,-\frac{M^2+2M\ell\omega-2E(M+\ell\omega)}{2\ell\omega\,E}\,.
\end{equation}
It is important to notice that $z_L(\frac{M}{2})\,=\,-1$ and $z_L(\frac{M}{2}+\ell\omega)\,=\,+1$.
We can rewrite these limits, for both signs of $\ell$,  as
\begin{equation}\label{I3}
\int\,dE\,\int\,dz\,\Theta(Q^2)\,=\,\int_{m}^{\frac{M}{2}}\,dE\,\int_{-1}^{1}\,dz
                                         +\int_{\frac{M}{2}}^{\frac{M}{2}+\ell\omega}dE\,\int_{z_L(E)}^1\,dz\,.
\end{equation}
We will treat these two terms separately.

Consider the first term in (\ref{I3}).  Because the limits of integration are not functions of $\ell$ we can
immediately do the sum over $\ell$ in Eq.\,(\ref{TG}).  The only things we need to do the sums are the relations
\begin{equation}\label{sum}
\sum_{\ell=-\infty}^{\infty}J_{\ell}(z)J_{\ell+n}(z)\,=\,J_n(0)\,,
\end{equation}
which follows from {\bf 9.1.75} in Abramowitz and Stegun\cite{AS}, the recurrsion relation for Bessel functions,
$\ell\,J_{\ell}(z)\,=\,\frac{z}{2}(J_{\ell+1}(z)+J_{\ell-1}(z))$, and $J_{-\ell}(z)\,=\,(-1)^{\ell}J_{\ell}(z)$.
The most important sum is
\begin{equation}\label{sum1}
\sum_{\ell=-\infty}^{\infty}J_{\ell}^2(D)\,=\,1
\end{equation}
because that replaces the $J_{\ell}^2(D)$ in the first line of (\ref{T2}) by unity.
After doing all the sums Eq.\,(\ref{T2}) becomes
\begin{eqnarray}
\sum_{\ell=-\infty}^{\infty}|T_{\ell}|^2\,&=&\,128\,G^2\,\big\{3P\cdot\,q(M^2+m^2)-4(P\cdot\,q)^2-2M^2m^2  \nonumber  \\
   &&\,\,\,\,\, +2\frac{e^2a^2}{(q\cdot\,k)^2}q_{x}^2\,P\cdot\,k(q\cdot\,k-P\cdot\,k)  \nonumber  \\
   &&\,\,\,\,\,\, -i\frac{e^2a^2}{q\cdot\,k}\epsilon(P,k,n_1,n_2)(2P\cdot\,k-2q\cdot\,k)\big\}  \label{Tl}
\end{eqnarray}
where there has been a lot of cancellation.
This can be easily integrated by hand
\begin{equation}\label{Gs}
%\Gamma\,=\,\Gamma^0\,+\,\frac{G^2e^2a^2M^3}{24\pi^3}\Big\{\frac{5}{3}-\ln\frac{M}{m}
%                           -\frac{2\omega}{M}\ln\frac{M}{m}+\frac{5}{2}\frac{\omega}{M}\Big\} \label{Gs}
\Gamma\,=\,\,\Gamma^0\,\Big(1+8\frac{e^2a^2}{M^2}\Big\{\frac{5}{3}-\ln\frac{M}{m}-\frac{2\omega}{M}\ln\frac{M}{m}
                           +\frac{5}{2}\frac{\omega}{M}\Big\}\Big)
\end{equation}
% THE SIGNS HAVE BEEN CHANGED ON THE OMEGA TERMS AS PER WAYNE
where again the first line of (\ref{Tl}) gives $\Gamma^0$.
Since $ea$ is much less than $1$\,MeV the
additional laser dependent terms are very small, smaller than the electron mass terms
which were neglected in (\ref{G0}).  The terms which are even further suppressed by the factor of $\omega/M$ come from
the $\epsilon$ term in (\ref{Tl}).

Now consider the correction from the second term in (\ref{I3}).
For $ea\sim10^{-4}$\,MeV
and $\omega\sim1\,{\rm eV}$,
the argument of the Bessel function,
$D$, varies with $E$ and $z$ from zero to $\sim\,1.5\times10^4$.  Bessel functions become very small once the index, $\ell$,
becomes greater than the argument
%So to save computer time and to avoid nasty underflow we should limit the integration
%by
so $\ell$ is limited by $\ell\,\le\,D$\,.
%\begin{equation}\label{theta}
%\theta\big(D-|\ell|\big)
%\end{equation}
In other words $\omega\ell$ will always be much less than $1$\,MeV and this correction will be small because the range
of the energy integration is small.  So define a function of the energy as
\begin{equation}\label{fE}
F(E,\omega,\ell)\,=\,\int\,dEE\beta\int_{z_L}^1\,dz\widetilde{|T_{\ell}|^2}
\end{equation}
where $\widetilde{|T_{\ell}|^2}$ is given by Eq.\,(\ref{T2}) without the prefactor of $128\,G^2$.
Then the correction to the width is given by
\begin{equation}\label{GC1}
\frac{\Gamma^C}{\Gamma^0}\,=\,\frac{8}{M^6}\sum_{\ell=-\infty}^{\infty}\big[F(E=\frac{M}{2}+\ell\omega,\,\omega,\,\ell)
                                                                       -F(E=\frac{M}{2},\,\omega,\,\ell)\big]\,.
\end{equation}
Now if we expand the first term in a Taylor series
\begin{equation}\label{Taylor}
F(E=\frac{M}{2}+\ell\omega,\,\omega,\,\ell)\,=\,F(E=\frac{M}{2},\,\omega,\,\ell)+\ell\omega\frac{dF}{dE}\rfloor_{E=\frac{M}{2}}
                                                           +\frac{\ell^2\omega^2}{2}\frac{d^2F}{dE^2}\rfloor_{E=\frac{M}{2}}
                                                           +\cdots
\end{equation}
then, since $z_L(\frac{M}{2})\,=\,-1$, we can again do the sum over $\ell$ before we do the integral.
We will approximate (\ref{GC1}) by keeping only the first nonzero term; since
\begin{equation}\label{dFdE}
\frac{dF}{dE}\rfloor_{E=\frac{M}{2}}\,=\,\frac{M}{2}\int_{-1}^1\,dz\widetilde{|T_{\ell}|^2}
\end{equation}
the correction, Eq.\,(\ref{GC1}), is
\begin{equation}
\frac{\Gamma^C}{\Gamma^0}\,=\,\frac{4\omega}{M^5}\int_{-1}^1\,dz\sum_{\ell=-\infty}^{\infty}\ell\widetilde{|T_{\ell}|^2}\,.
\end{equation}

Now in Eq.\,(\ref{T2}) the only nonzero sums come from lines 3, 4, 7, 9, 10, and 12 and after a bit of work we get
\begin{equation}\label{GCT}
\frac{\Gamma^C}{\Gamma^0}\,=\,8\frac{e^2a^2}{M^2}\Big\{\big(1+3\frac{\omega}{M}+4\frac{\omega^2}{M^2}\big)\ln\frac{M}{m}
                                                      -1-4\frac{\omega}{M}-2\frac{\omega^2}{M^2}\Big\}\,.
\end{equation}
These terms are small but not necessarily smaller than those in (\ref{Gs}).  In fact the largest term cancels and
the total effect of the laser, the sum of (\ref{Gs}) and (\ref{GCT}), is to change the width from $\Gamma^0$ to
\begin{equation}\label{Soln}
\Gamma\,=\,\Gamma^0\Big\{1+8\frac{e^2a^2}{M^2}\Big[\big(\frac{\omega}{M}+4\frac{\omega^2}{M^2}\big)\ln\frac{M}{m}
+\frac{2}{3}-\frac{3}{2}\frac{\omega}{M}-2\frac{\omega^2}{M^2}\Big]\Big\}\,.
\end{equation}

Let's be clear about what has been done.  Eq.\,(\ref{Gs}) is an exact solution for the part of the phase space
where the integral over the electron energy is between $m$ and $\frac{M}{2}$.  Eq.\,(\ref{GCT}) is an approximation for
the part of the phase space where the integral is between $\frac{M}{2}$ and $\frac{M}{2}+\ell\omega$, in that it
is the first nonzero term in an expansion in powers of $\ell\omega$.  As a consistancy check
we have calculated the contribution to the second
nonzero term in the
expansion (the second derivative term in (\ref{Taylor}))
from the first line of (\ref{T2}) (which should be the largest contribution) and found
\begin{equation}
\frac{\Gamma^C}{\Gamma^0}\,\sim\,32\frac{e^2a^2m^2}{M^4}[-1+\ln\frac{M}{m}]
\end{equation}
which is the same as zero since we did not keep terms $\sim\frac{m}{M}$ in (\ref{Gs}) or (\ref{GCT}).

\section{Comparison with LLB}

We have assumed the laser radiation is circularly polarized because, as is well known\cite{Todd}, that case is easier.
For linear polarization the wavefunction in Eq.\,(\ref{wf}) is multiplied by an extra factor
\begin{equation}\label{wf1}
e^{i\frac{e^2a^2}{8\,p\cdot\,k}\sin(2k\cdot\,x)}
\end{equation}
and the second term in the definition of $q^{\mu}$, Eq.\,(\ref{q}), has an extra $\frac{1}{2}$.
Because of this term the generating function for Bessel functions must be used twice and the square of the
matrix element involves two sums over Bessel functions.  LLB use linear polarization but approximate the wavefunction
by omitting this term thus avoiding the double sum.  Dropping this term is not the same as doing circular polarization
because the vector potential, $A^{\mu}$, is different.  The effect of this approximation on the muon lifetime is unknown
but it seems unlikely this could explain the huge difference in our results.
Given their definition of the wavefunction we agree with their expression for the S matrix.
We do not, however, agree with their expression for the partial width.  We believe their
$W_{\ell}\,=\,\frac{G^2}{96\pi^4}\int\,dE\cdots$ should be $W_{\ell}\,=\,\frac{G^2}{48\pi^4\,M}\int\,dEE\cdots$.
($W_{\ell}$ is what we call $\Gamma_{\ell}$.)  Otherwise their $\Gamma^0$ would be too big by a factor of $\frac{5}{3}$.
This is probably a misprint.  If not, it would partially explain their small lifetimes.

LLB find a shorter lifetime which means they find a larger width.  On the other hand they
don't get much contribution from what we call negative $\ell$.
(We have defined $\ell$ differently -- what we call negative $\ell$ LLB call positive.)
If we consider only positive $\ell$
then, for example, Eq.(\ref{sum1})
is replaced by
\begin{equation}
\sum_{\ell=0}^{\infty}J_{\ell}^2(z)\,=\,\frac{1}{2}(1+J_0^2(z))
\end{equation}
and $\Gamma^0$, the contribution from the first line of (\ref{T2}),
is replaced by approximately $\frac{1}{2}\Gamma^0$.  {\it But this is a change in the wrong direction.}
LLB get a larger width {\it despite} having little contribution from negative $\ell$.
As we have seen the other terms
in (\ref{T2}) go as $e^2a^2/M^2$.
%For us the negative $\ell$ give the other $\frac{1}{2}$
%\begin{equation}
%\sum_{\ell=-1}^{-\infty}\,J_{\ell}^2(z)\,=\,\frac{1}{2}(1-J_0^2(z))
%\end{equation}
Furthermore a {\it longer} lifetime might be easier to understand as a kind of Zeno effect; the interactions with
the laser photons make the decay keep starting over.  But that is not what LLB find.

We disagree with LLB on the lower limit of the electron energy. We get the lower limit to be the (effective)
electron mass, $m$, they get $m+\ell\,\omega$.  Again it is hard to believe that makes much difference.

So about the only place left to look for a difference is the numerical integrations.  To do these requires
definite values for $ea$ and $\omega$.  We used $ea\,=\,1.69\times10^{-4}$\,MeV and $\omega\,=\,1.17$\,eV
(the Nd:YAG laser of LLB) and integrated (\ref{T2}) using the limits (\ref{lim1}) or (\ref{lim2}).
A few results are shown in the table
\begin{center}
\begin{tabular}{|c|c|}  \hline
$\ell$ & $\frac{\Gamma_{\ell}}{\Gamma^0}$  \\ \hline
0 & 3.463 $10^{-3}$  \\
1 & 3.455 $10^{-3}$ \\
10 & 3.427 $10^{-3}$ \\
100 & 1.921 $10^{-3}$ \\
200 & 6.940 $10^{-4}$  \\
500 & 7.391 $10^{-5}$  \\
1000 & 1.013 $10^{-5}$  \\
5000 & 7.866 $10^{-8}$  \\  \hline
\end{tabular}
\end{center}
We used routines for the Bessel functions from Numerical Recipes\cite{numrec}
and the integration routine VEGAS\cite{vegas}.
The partial widths for negative $\ell$ were indistingushable from those of positive $\ell$ for a given $|\ell|$.
The integration was sufficiently fast that we could do each of the $\Gamma_{\ell}$ for $\ell$ up to $500$
with the result
\begin{equation}\label{sum500}
\frac{1}{\Gamma^0}\sum_{\ell=-500}^{500}\,\Gamma_{\ell}\,=\,0.96
\end{equation}
where we estimate the error from the numerical integration to be less than $0.01$.
If we assume the $\ell$ dependence of $\Gamma_{\ell}$ is linear from $\ell\,=\,500$ to $1000$
we get another contribution to (\ref{sum500}) of $0.04$;
if we make the same assumption for $\ell$ from $1000$ to $5000$ we get another $0.04$.
These are surely overestimates because the $\ell$ dependence must fall faster than linear.
We could do a better job for $|\ell|\,>\,500$ but these are sufficient to show the magnitude of the contribution
to the total width that could be expected from higher $\ell$.
Thus the conclusion is that the total width cannot be very different than $\Gamma^0$, in agreement with our
more precise arguments above.

\section{Conclusions}

We have considered muon decay in the electromagnetic field of a laser. Our discussion is entirely analytic
with the only approximation a Taylor series expansion of the squared matrix element for the region of
electron energy between $\frac{M}{2}$ and $\frac{M}{2}+\ell\omega$, a distance of less than $10^{-2}$\,MeV.
We find the effect of a laser on muon decay is very small, of order $e^2a^2/M^2$, $e^2a^2\omega/M^3$, or
$e^2a^2\omega^2/M^4$,
where $ea\,\sim\,2\times10^{-4}$\,MeV, $\omega\,\sim\,1$\,eV
and $M$ is the muon mass, $105.66$\,MeV.
Given the dimensions of the Fermi coupling constant the decay width must have five powers of energy.
Once we find the coefficient of the $M^5$ term is the same as in the absence of the laser we know the effect
of the laser is very slight because the only other energies in the problem are $ea$, $\omega$, and $m$.
%Our full result is given by Eq.\,(\ref{Soln}).
We dropped the electron mass where possible so there are surely corrections of order $e^2a^2m^2/M^4$
which are numerically larger than some of the terms we included.
But all of these corrections are tiny, smaller than the known corrections of order $m^2/M^2$.
Our result is given by Eq.\,(\ref{Soln}).

%We could ask what $\ell\,=\,\infty$ means in practice and the answer is that because $\ell\,\le\,D$ it means
%$\ell\,\approx\,10^6$.  I don't know if a laser has $10^6$ photons in a pulse or how much of the pulse can be concentrated
%on a single muon but I don't think that is relevant because the sum over $\ell$ {\it does not mean there are $\ell$
%photons being absorbed or emitted}.  If that were true then the width in the absence of photons, $\Gamma^0$, would
%have to equal $\Gamma_{\ell}$ with $\ell\,=\,0$.  In fact the opposite is approximately true, $\Gamma^0$ is the sum over all
%$\ell$ of $\Gamma_{\ell}$.  The sum over $\ell$ comes from converting the Volkov wavefunction to a sum over Bessel functions
%using the generating function -- it is a mathematical statement not a physical statement.  We could calculate the probability
%for absorbing or emitting $\ell$ photons and then sum over $\ell$; that is what we do when we calculate radiative
%corrections.  That is {\it not} what we are doing here.

\section*{Acknowledgments}
DAD was supported in part by the U. S. Department of Energy under grant No. DE-FG03-93ER40757.
WWR was supported in part by the National Science Foundation under Grant PHY-0555544.

\end{document}